\documentclass[12pt,a4paper]{article}
\usepackage{graphicx}
\begin{document}
 
\title{\bf Dynamics of competition between collectivity and noise in the stock market}
 
\author{S. Dro\.zd\.z$^{1,2}$, F. Gr\"ummer${^1}$, F. Ruf$^{3}$ and J. Speth$^{1}$\\
\\
$^{1}$Institut f\"ur Kernphysik, Forschungszentrum J\"ulich,\\
D-52425 J\"ulich, Germany \\
$^{2}$ Institute of Nuclear Physics, PL-31-342 Krak\'ow, Poland\\
$^{3}$ WestLB International S.A., 32-34 bd Grande-Duch. Charl.,\\
L-2014 Luxembourg}
\date{\today}
\maketitle

\begin{abstract}
Detailed study of the financial empirical correlation matrix
of the 30 companies comprised by DAX within the period of the last 11 years,
using the time-window of 30 trading days, is presented.
This allows to clearly identify a nontrivial time-dependence of the resulting
correlations. In addition, as a rule, the draw downs are always accompanied
by a sizable separation of one strong collective eigenstate 
of the correlation matrix which, at the same time, 
reduces the variance of the noise states. The opposite applies to draw ups.
In this case the dynamics spreads more uniformly over the eigenstates which
results in an increase of the total information entropy.

\end{abstract}
 
 
\bigskip
 
\newpage

Studying correlations among various financial assets is of great interest
both for practical as well as for fundamental reasons. Practical aspects 
relate for instance to the theory of optimal portfolios and risk
management~\cite{Mark}.
The fundamental interest, on the other hand, results from the fact that
such study may shed more light on the universal aspects of evolution
of complex systems.  
Recent study~\cite{Lalo,Pler} of the related problems in the context 
of the stock market show that majority of eigenvalues in the spectrum of the
correlation matrix agree very well with the universal predictions of
random matrix theory. Locations of some of the eigenvalues differ however
from these predictions and thus identify certain system-specific, non-random 
properties of the system. The corresponding eigenvalues thus carry most of the
information about the system. The above studies have however a global in time 
character and do not account for a possible change of correlations on shorter
time-scales.

For a set of N assets labelled with $i$ and represented by the time-series 
of price changes $\delta x_i(t)$ of length $T$ one forms a $N \times T$ 
rectangular matrix $\bf M$. Then, the correlation matrix is defined 
as 
\begin{equation}
{\bf C} = {1\over T} {\bf M} {\bf M}^{\bf T}.
\label{eq:C}
\end{equation}
In order not to artificially reduce the rank of this matrix, $T$ needs to
be at least equal to $N$. This sets the lowest limit on a time-window which
can be used to study the time-dependence of correlations.   
One of the best examples where $T$ can be set relatively low, and thus
efficiently allow to get some time-dependent picture of correlations versus 
the global market index, is provided by DAX (Deutsche Aktienindex).  
It represents a matured, relatively independent market, whose behavior is well 
reflected by $N=30$ companies defining this index. During the period 
studied here it displays the whole variety of behaviors like stagnancy, booms,
including a pattern of self-similar log-periodic structures~\cite{Droz2}, 
and crashes. 

The present study is based on daily variation of all $N=30$ assets of the DAX
during the years 1988-1999. When calculating the covariance matrix the average
value of the price changes is subtracted off and then their values rescaled
so that $\sigma^2 = \langle \delta x_i^2 \rangle = 1$.  
The time-interval $T$ is set to 30 and continuously moved over the whole 
period.  

That the character of correlations may 
significantly vary in time is indicated in Fig.~1 which displays some typical  
distributions of matrix elements of $\bf C$ in three different cases: 
(i) an average over all time-intervals of length $T=30$, 
(ii) for a single $T=30$ time-interval which ends on November 25, 1997,
(iii) for a single $T=30$ time-interval which ends on April 7, 1998.
Clearly, in all the cases the distributions are Gaussian-like but their 
variance and location is significantly different.  
In fact, a distribution of this type about prescribes the structure 
of the corresponding eigenspectrum. The point is that to a first approximation
any of such matrices can be represented as 
\begin{equation}
{\bf C} = {\bf G} + \gamma {\bf U},
\label{eq:G}
\end{equation}
where $\bf G$ is a Gaussian centered at zero and $\bf U$ is a matrix whose
all entries are equal to unity and $0 \leq \gamma \leq 1$.   
The rank of matrix $\bf U$ is one, therefore the second term alone in the above
equation generates only one nonzero eigenvalue of magnitude $\gamma$.
As the expansion coefficients of this particular state are all equal
this assigns a maximum of collectivity to such a state.
If $\gamma$ is significantly larger than zero the structure of the matrix 
${\bf U}$ is dominated by the second term in (\ref{eq:G}) 
and an anticipated result is one collective state with large eigenvalue. 
Since in this case ${\bf G}$ can be considered only a 'noise' correction 
to $\gamma {\bf U}$ all the other states are connected with small eigenvalues. 
The above provides an alternative potential mechanism for emergence of
collectivity out of randomness to the one taking place in finite interacting
Fermi systems. There, a reduction of dimensionality of a leading component
in the Hamiltonian matrix is associated with appearance of more 
substantial tails in the distribution of large matrix elements 
as compared to an ensemble of random matrices~\cite{Droz2}.    

A relatively small number of stocks comprised by DAX makes somewhat difficult 
a full statistical analysis of the 'noise' term. Its properties, analogous
to the predictions of the Gaussian orthogonal ensemble (GOE)~\cite{Wign} 
of random matrices, are however already well established 
in the recent literature~\cite{Lalo,Pler}.
In the present case instead, one can trace a possible non stationarity in the
location of eigenvalues with a comparatively good time-resolution.
The corresponding central result of our paper is displayed in Fig.~2.  
As it is clearly seen, the draw ups and the draw downs of the global index (DAX),
respectively, are governed by dynamics of significantly distinct nature.  
The draw downs are always dominated by one strongly
collective eigenstate with large eigenvalue. 
Such a state thus exhausts a dominant fraction of the total portfolio variance 
\begin{equation}
\sigma_P = \sum^N_{i,j} p_i C_{ij} p_j,
\label{eq:sigma}
\end{equation}
where $p_i$ expresses a relative amount of capital invested in the asset $i$,
and $C_{ij}$ are the entries of the covariance matrix $\bf C$~\cite{Mark,Bouc}.
(This becomes obvious by transforming $\bf C$ to its eigenbasis.)
The more dramatic the fall is the more pronounced
is this effect. At the same time, by conservation of the trace of ${\bf C}$, 
the remaining eigenvalues (representing some less risky portfolios) are
compressed in the region close to zero. In a formal sense, this effect
is reminiscent of the slaving principle of synergetics~\cite{Hake}:
one state takes the entire collectivity by enslaving all the others. 

The opposite applies to draw ups.
Their onset is always accompanied by a sizable restructuring of 
eigenvalues towards a more uniform distribution. The related principal effect
is that the largest eigenvalue moves down which is compensated by
a simultaneous elevation of lower eigenvalues. At one instant of time 
(mid 1996), which marks the beginning of the long-term boom, the two largest
eigenvalues become almost degenerate. Based on these results a general 
statement that an increase on the market involves more competition 
than a decrease seems appropriate since in the former case 
the total variance is more democratically distributed among eigenstates 
of the correlation matrix. In other words, the increase on the stock market,
at least in the presently studied case,
never involves parallel uniform increase of prices of all the
participating companies as it happens during decreases.

Such a conclusion is also indicated by the information entropy:     
\begin{equation}
I_k= \sum_{l=1}^{30} -(u_{kl})^2 \ln (u_{kl})^2,
\label{eq:Ient}
\end{equation}
where $u_{kl}$ (here $l=1,...,30$) are the components of eigenvector $k$.
Its GOE limit~\cite{Izra} is
\begin{equation}
I^{GOE} = \psi(N/2 +1) - \psi(3/2)
\label{eq:IGOE}
\end{equation}
where $\psi$ is the digamma function and $N$, in the present study, 
corresponds to the number of stocks. 
For $N=30$ we thus have $I_k^{GOE} \approx 2.67$
and for the uniformly distributed components $(u_{kl}^2=1/30)$ $I_k^{uni}=3.4$.
These two limits are to be related to the information entropies displayed
in the lower panel of Fig.~2. As one can see, it happens only during decreases
that the information entropy approaches the limit of uniform distribution
for the upper, most collective state. Otherwise this state becomes somewhat 
more localized. The information entropy of the 'noise' states on average
about agrees with $I^{GOE}$, though, a more careful inspection shows systematic
and consistent deviations, going in opposite directions during increases 
and decreases, respectively. 

In quantitative terms this effect 
can easily be deduced by looking at the total information entropy
\begin{equation}
I_{tot} = \sum_{k=1}^{30} I_k 
\label{eq:Itot}
\end{equation}
shown in Fig.~3. On average it reveals a visible tendency of
moving in opposite direction relative to the information entropy
$I_1$ of the most collective state, even though $I_1$ is included in $I_{tot}$.
This result is very interesting and even intriguing in itself.     
The market draw ups are accompanied by increase of the total information entropy
while draw downs are associated with its decrease. 
At first glance such a behavior and, especially, the entropy decreases 
accompanying such turmoils as crashes may look somewhat embarrassing.
A possible reason for this effect might be the fact that prices 
and related quantities reflect only a part of the market world. 
There is also an environment with which any 
market constantly interacts and which easily may absorb a corresponding portion
of entropy. Actually, the turmoils accompanying crashes are visible more 
in the market environment than in the genuine market parameters.         

Since the structure of the covariance matrix is influenced by 
measurement noise more for short time-series than for the long ones, 
a question which needs to be answered is to what extent our conclusions 
are stable with respect to the length $T$ of the price time-series.
Of course, the specific values of the quantities considered do depend on $T$
but the main effect of increasing it is to smear them out in time. 
The global tendencies, of interest for the present paper, remain however
unchanged. An example is shown in Fig.~4  which displays the structure of
eigenspectrum of the covariance matrix for several values of $T$. 
 
In summary, the present study discloses several interesting novel facts
about the dynamics of financial evolution.
These empirical facts, interpreted in terms of the coexistence of collectivity
and noise in correlations among the financial assets, provide
arguments for distinct nature of the mechanism governing financial increases
and decreases, respectively, even though such correlations on average are
largely compatible with the random matrix theory predictions.
The structure of eigenspectrum of the correlation matrix and the information
entropy arguments point to increases as those phenomena which internally 
involve more diversity and competition as compared to decreases.
It seems likely that such characteristics may apply to the dynamics 
of evolution of other complex systems as well.

\newpage

\begin{figure}[ht]
\begin{center}
\leavevmode
\includegraphics[width=12cm]{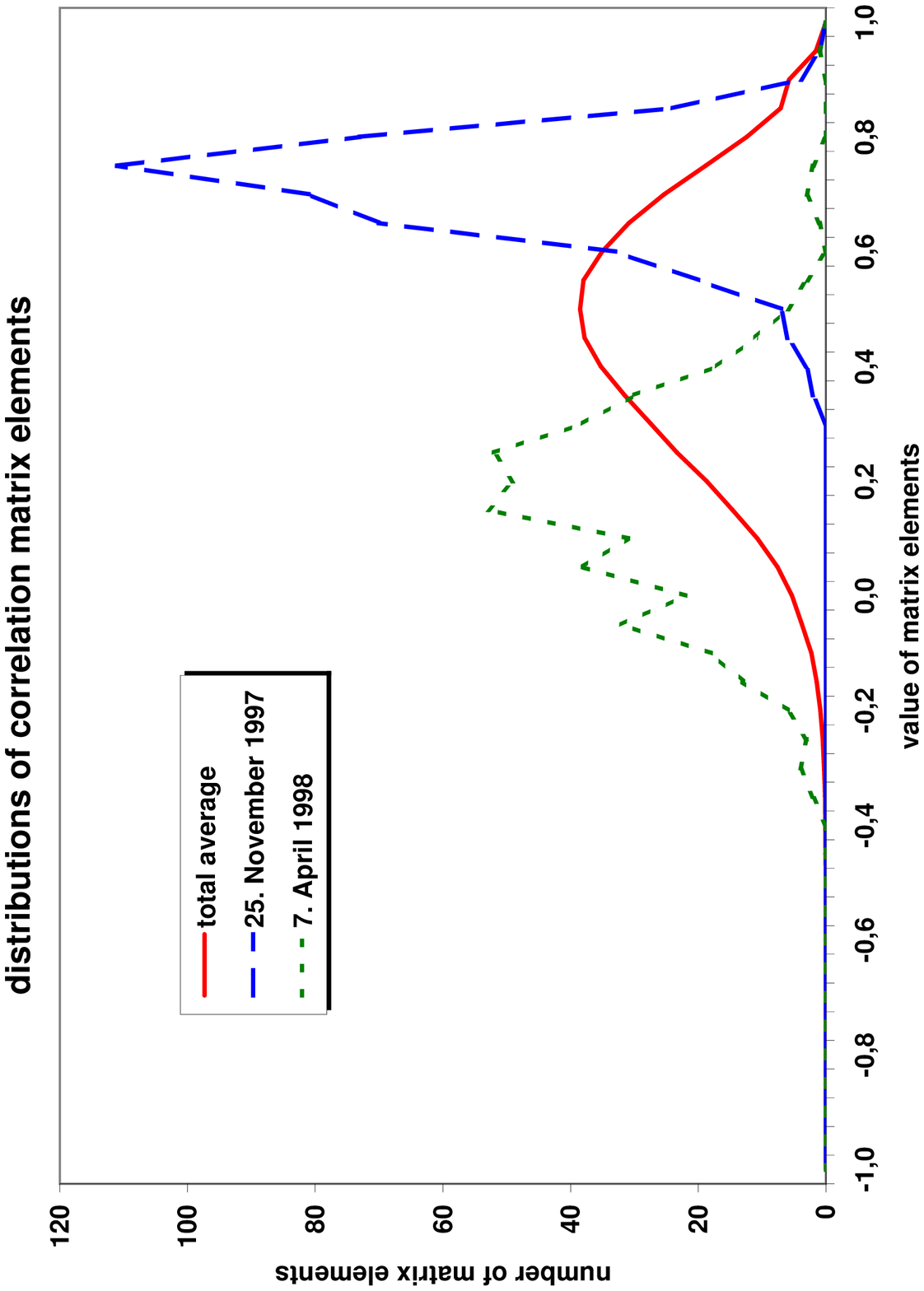}
\end{center}
\caption{\it Distributions of matrix elements of the
correlation matrix $\bf C$ calculated from daily price variation of all
$N=30$ companies comprised by DAX in three different cases: 
(i) an average over all time-intervals of length $T=30$ during the period
1988-1999. 
(ii) for a single $T=30$ time-interval which ends on November 25, 1997,
(iii) for a single $T=30$ time-interval which ends on April 7, 1998.} \label{fig1}
\end{figure}

\begin{figure}[ht]
\begin{center}
\leavevmode
\includegraphics[width=14cm]{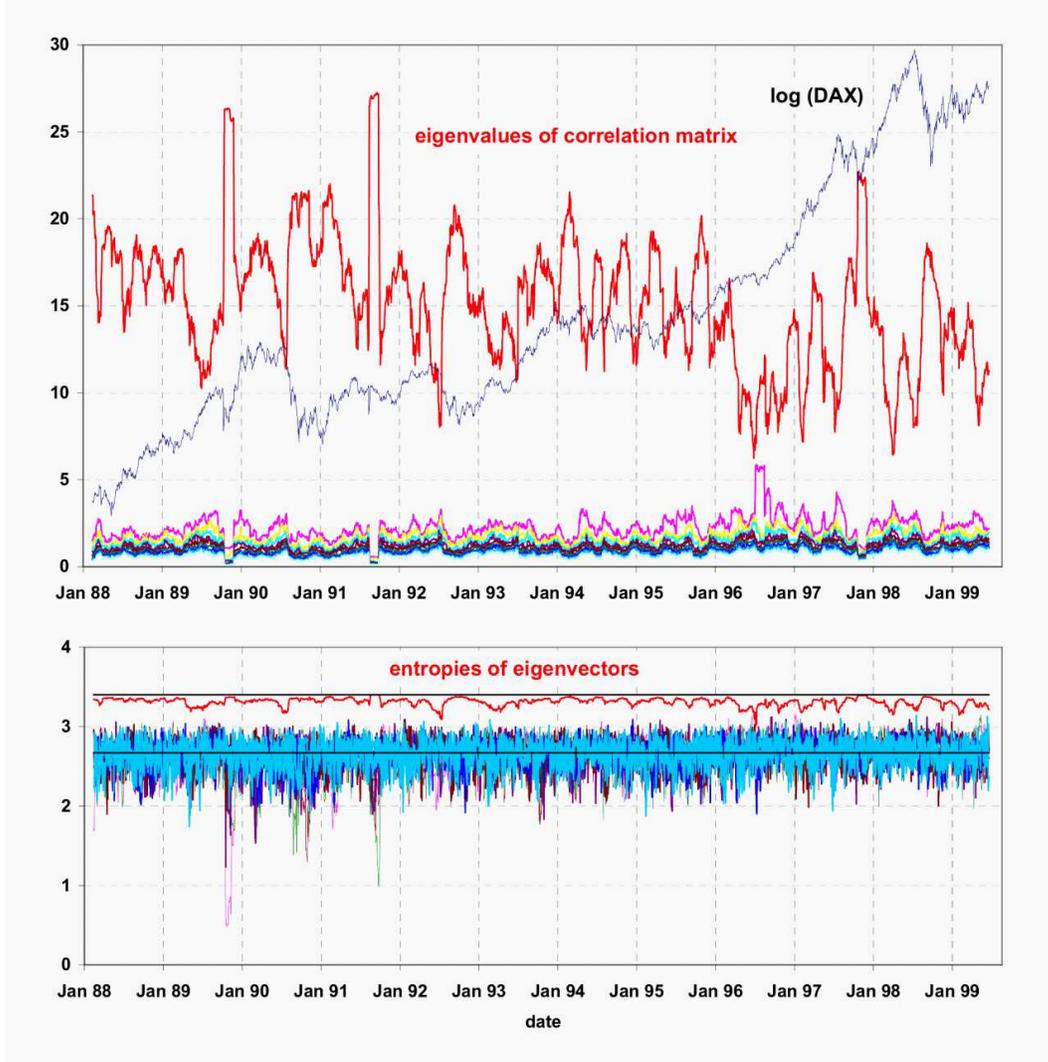}
\end{center}
\caption{\it Time-dependence of the 10 largest eigenvalues (upper panel) and of the
related spectrum of information entropies (lower panel) corresponding  
to the DAX correlation matrix $\bf C$ calculated from the time-series  
of daily price changes in the interval of $T=30$ past days, during the 
years 1988-1999.
The DAX time-variation (represented by its logarithm) during the same period
is displayed in the upper panel. The two solid horizontal lines in the lower
panel indicate the two reference limit values, $I^{GOE} = 2.67$ 
and $I^{uni}=3.4$, of the information entropy.} \label{fig2}
\end{figure}

\begin{figure}[ht]
\begin{center}
\leavevmode
\includegraphics[width=12cm]{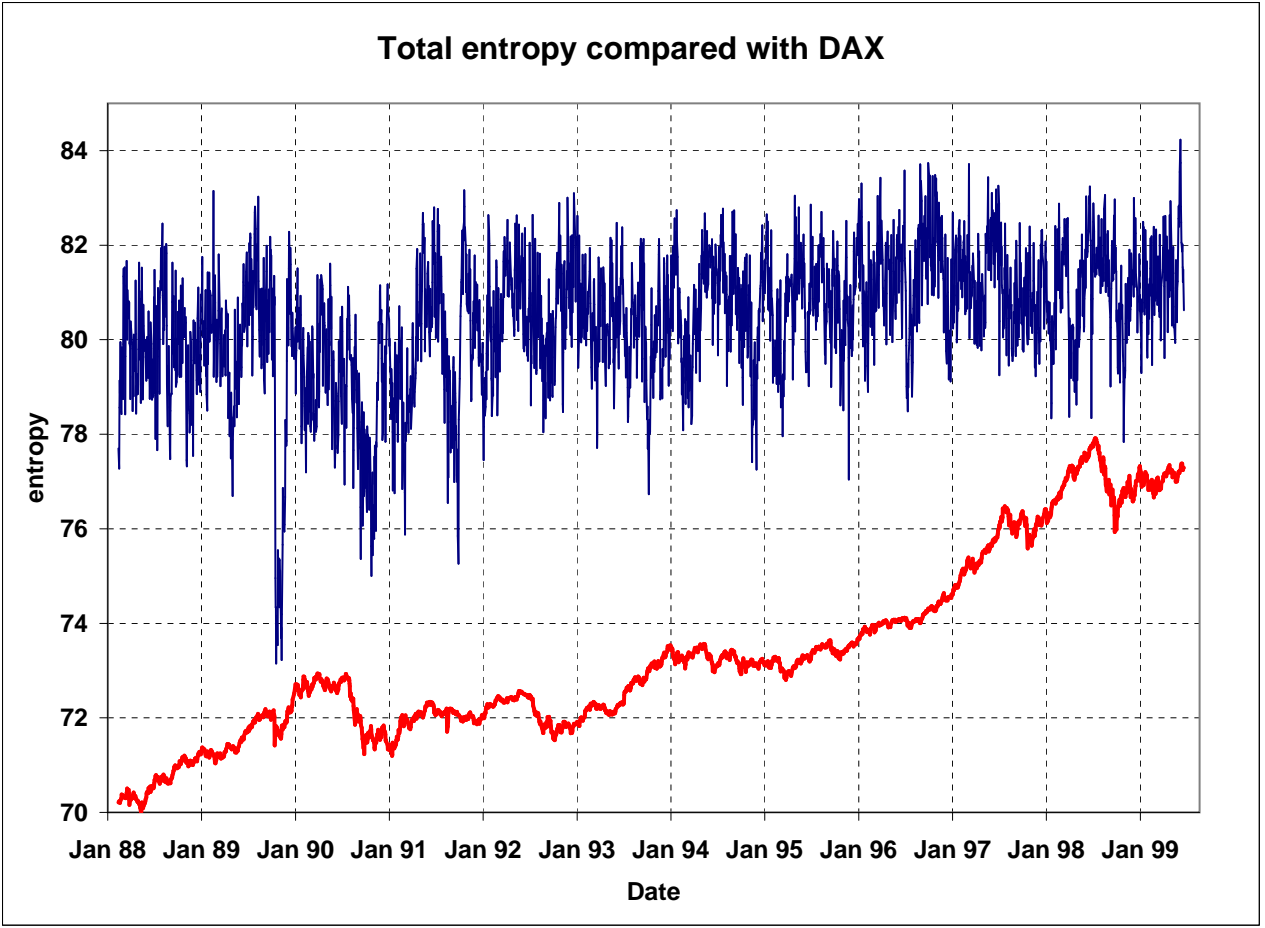}
\end{center}
\caption{\it Time-dependence of the total information entropy $I_{tot}$
versus DAX.} \label{fig3}
\end{figure}

\begin{figure}[ht]
\begin{center}
\leavevmode
\includegraphics[width=12cm]{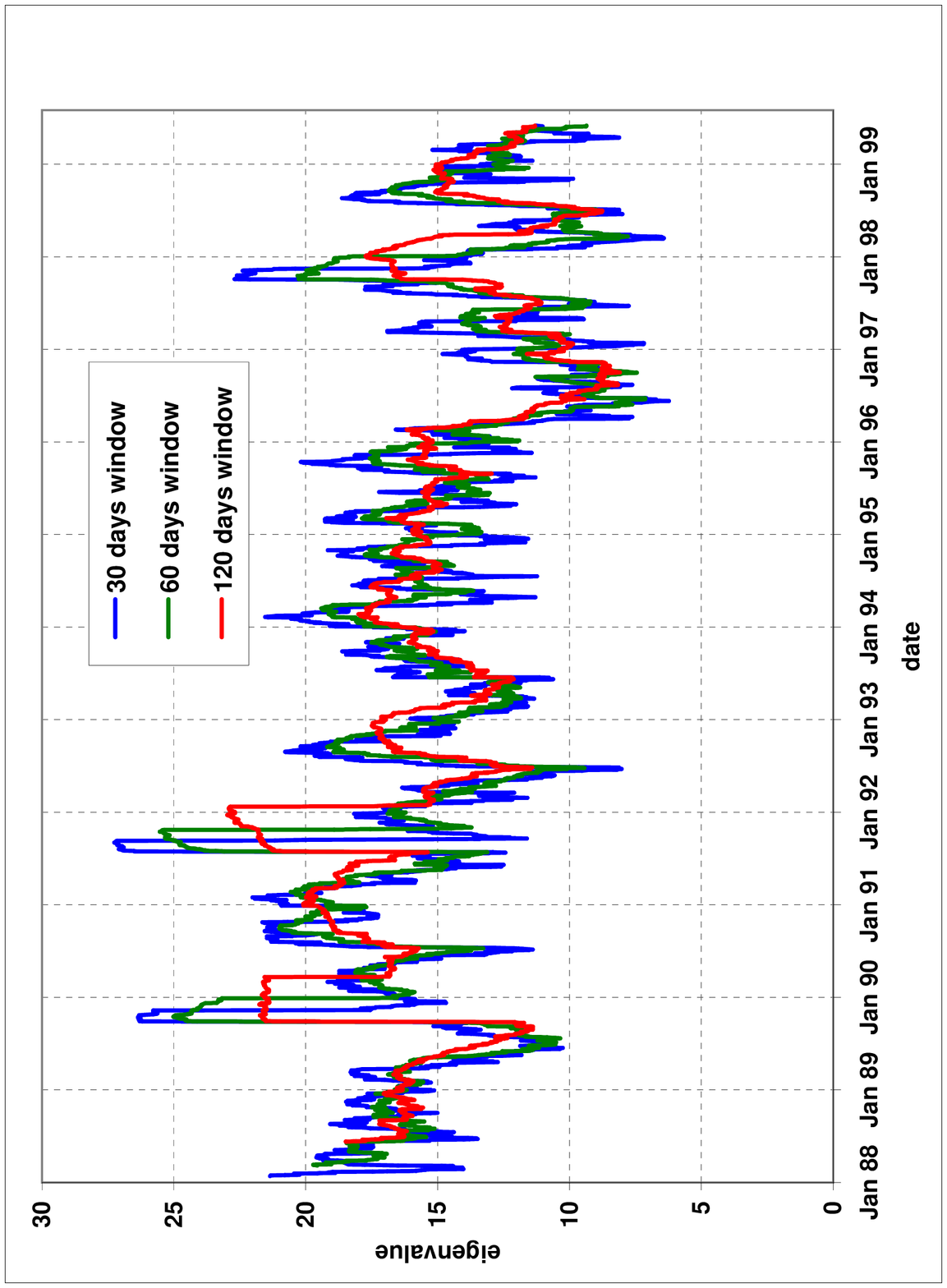}
\end{center}
\caption{\it Time-dependence of eigenspectra (as in Fig.~2) for 
$T=30$, 60 and 120 days.} \label{fig4}
\end{figure}


\bigskip 


\newpage

\begin{thebibliography}{99}

\bibitem{Mark} H.~Markowitz, {\it Portfolio Selection: 
Efficient Diversification of Investments} 
(J.~Wiley and Sons, New York, 1959);\\
E.J.~Elton and M.J.~Gruber, 
{\it Modern Portfolio Theory and Investment Analysis}
(J.~Wiley and Sons, New York, 1995)
\bibitem{Lalo} L.~Laloux, P.~Cizeau, J-.P~Bouchaud and M.~Potters,
Phys. Rev. Lett. {\bf 83}, 1467(1999)
\bibitem{Pler} V.~Plerou, P.~Gopikrishnan, B.~Rosenow, L.A.~Nunes Amaral
and H.E.~Stanley, Phys. Rev. Lett. {\bf 83}, 1471(1999)
\bibitem{Meht} M.L.~Mehta, Random Matrices (Academic Press, Boston, 1999)
\bibitem{Bouc} J.M.~Bouchaud and M.~Potters, {\it Theory of Financial Risk},
(Al\'ea-Saclay, Eyrolles, Paris, 1997) (in Franch)
\bibitem{Droz1} S.~Dro\.zd\.z, F.~Ruf, J.~Speth and M.~W\'ojcik,
Eur. Phys. J. {\bf B10}, 589(1999)
\bibitem{Droz2} S.~Dro\.zd\.z, S.~Nishizaki, J.~Speth and M.~W\'ojcik,
Phys. Rev. {\bf E57}, 4016(1998) 
\bibitem{Wign} E.P.~Wigner, Ann. Math. {\bf 53}, 36(1951)
\bibitem{Hake}  H.~Haken, {\it Advanced Synergetics} (Berlin, Speringer,
1987);\\  {\it Information and Selforganization} (Berlin, Springer, 1988)
\bibitem{Izra} F.M.~Izrailev, Phys. Rep. {\bf 196}, 299(1990)
 




\end{thebibliography}
\end{document}